\documentclass[11pt]{article}
\usepackage{latexsym}  
\usepackage{amssymb}
\usepackage{graphicx}
\usepackage{amsmath}

\begin{document}


\begin{center}

{\Large\bf Tantalizing New Physics from the Cosmic Purview } 

\vspace{4mm}

{\bf Fulvio Melia$^1$}

{Department of Physics, The Applied Math Program, and Department of Astronomy,
The University of Arizona, AZ 85721, USA. E-mail: fmelia@email.arizona.edu}

$^1$John Woodruff Simpson Fellow.

\date{\today}
\end{center}

\vspace{3mm}

\begin{abstract}
The emergence of a highly improbable coincidence in
cosmological observations speaks to a remarkably simple cosmic expansion.
Compelling evidence now suggests that the Universe's gravitational horizon,
coincident with the better known Hubble sphere, has a radius improbably equal to
the distance light could have travelled since the Big Bang. The confirmation of
this unexpected result would undoubtedly herald the influence of new physics,
yet appears to be unavoidable after a recent demonstration that the
Friedmann-Lema\^itre-Robertson-Walker metric is valid only for the so-called
zero active mass equation of state. As it turns out, a cosmic fluid with this
property automatically produces the aforementioned equality, leaving little
room for a cosmological constant. The alternative---a dynamical dark energy---would
suggest an extension to the standard model of particle physics, and a serious
re-evaluation of the Universe's early history.
\end{abstract} 

PCAC numbers:  04.20.Jb, 95.30.Sf, 98.80.Es, 98.80.Jk

\vspace{3mm}

\noindent{\large\bf 1  \  Introduction}\\ 
As cosmological observations stretch our view of the Universe to
progressively higher redshifts, allowing us to probe physics at the
earliest moments after the Big Bang, general relativity (GR) reminds
us to be wary of limits to our exploration arising from the possible
influence of various horizons. In cosmology, the term `horizon' has
been used to characterize the maximum distance particles could have
traveled since the beginning---the observer's `particle horizon'---or
an imaginary surface that forever separates spacetime events that
are causally connected to each other from those that are not---the
`event horizon' \cite{Rindler:1956}. Several other definitions have
a more customized application, such as the `acoustic horizon',
characterizing the maximum comoving distance traveled by sound
waves prior to recombination, giving rise to the sub-degree
scale anisotropies seen in the cosmic microwave background (CMB).

But it is another type of horizon, less familiar than these,
that will command much of our attention in this paper. Known as an
`apparent horizon' in GR, this membrane may emerge
in general, spherically or non-spherically symmetric spacetimes,
subdividing congruences of ingoing and outgoing null geodesics from
a given compact region. In the case of spherical geometry,
these are just the ingoing and outgoing radial null geodesics
from a two-sphere of symmetry \cite{Ben-Dov:2007,Faraoni:2011,Bengtsson:2011,Faraoni:2015}.

Indeed, for a spherically-symmetric spacetime, such as the Friedmann-Lema\^itre-Robertson-Walker
(FLRW) metric used extensively in modern cosmology, the apparent horizon may be
understood more readily when characterized as a {\sl gravitational} horizon, with
a radius $R_{\rm h}$ that---in this context---coincides with the size of the better
known Hubble sphere \cite{Melia:2018a}. In fact, we now know that the Hubble radius
owes its existence to these gravitational effects. In retrospect, the role
played by the $R_{\rm h}$ horizon in our interpretation of the data---about which we shall
have much more to say later in this paper---should have been obvious many decades ago,
given that a form of $R_{\rm h}$ already appeared in the early 1900's in de Sitter's
\cite{deSitter:1917} first publication of his now famous solution to Einstein's
equations. No doubt, the popularization of comoving coordinates by Friedmann in
the 1920's \cite{Friedmann:1923} diverted attention away from $R_{\rm h}$, which
only appears in the metric when its coefficients are written in terms of
proper---rather than comoving---coordinates (see, e.g.,
refs.~\cite{Melia:2007,MeliaAbdelqader:2009}). An important feature of
the apparent (gravitational) horizon in cosmology is that its radius
$R_{\rm h}$ is time-dependent. It is not static, like an event horizon
surrounding a Schwarzschild black hole. The Universe's apparent horizon
may or may not turn into a true event horizon, depending on the equation
of state (i.e., pressure $p$ versus energy density $\rho$) in the cosmic fluid,
which determines the future history of the cosmic expansion.

It is not difficult to understand why an apparent (gravitational) horizon
emerges in FLRW, by simply comparing the two sets of metric coefficients
written, respectively, in terms of the comoving and proper coordinates
\cite{Melia:2018a}. The high degree of symmetry in this metric negates
any gravitational influence on the interior of a spherical shell from
the rest of the Universe outside, a clear dichotomy enabled by the
Birkhoff theorem and its corollary \cite{Birkhoff:1923,Melia:2007,Melia:2018a}.
Not surprisingly, then, $R_{\rm h}$ is simply given by the familiar
Schwarzschild form
\begin{equation}
R_{\rm h}={2GM\over c^2}\;,
\end{equation}
where $M$ is the {\it proper} mass contained within a sphere of proper
radius $R_{\rm h}$, i.e.,
\begin{equation}
M\equiv {4\pi\over 3}R_{\rm h}^3\,{\rho\over c^2}\;,
\end{equation}
in terms of the energy density $\rho$ in the cosmic fluid.
This mass is sometimes referred to as the Misner-Sharp mass
\cite{Misner:1964}, since it appeared in the pioneering work of
Misner and Sharp in their consideration of a general relativisitc
spherical collapse, and sometimes as the Misner-Sharp-Hernandez mass,
for the subsequent analysis carried out by these authors \cite{Hernandez:1966}.
As noted, however, one should not confuse the physical meaning of this
$R_{\rm h}$ in the cosmological context with that of a black hole since,
for these objects, the apparent horizon is static, and therefore an event
horizon, though this is not yet necessarily the case for the cosmos.

This somewhat pedagogical introduction to the mass $M$ and its impact on
$R_{\rm h}$ may not adequately convey the degree to which these definitions
have been studied in the past. For example, Equation~(1) is well constrained
by the fact that only the Misner-Sharp-Hernandez designation for $M$ is consistent
with the metric coefficient $g_{rr}$ in spherically-symmetric spacetimes for
gravitational expansion or collapse. In spite of the fact that it can be
quite difficult to identify the physical mass-energy for non-asymptotically
flat geometries in GR \cite{Faraoni:2015}, other possible
definitions, such as the Hawking-Hayward quasilocal mass \cite{Prain:2016},
are merely restatements of the Misner-Sharp-Hernandez construct.

Thus, although our definition of $R_{\rm h}$ appears to be overly simplistic,
it is actually fully consistent with the already understood
meaning of an apparent horizon in general relativity, including its relevance
to black-hole systems \cite{Melia:2018a}. More to the point of this paper,
the FLRW metric is always spherically symmetric, so the general definition
of an apparent horizon reduces exactly to Equation~(1) \cite{Faraoni:2011,Faraoni:2015}
and, for this reason, we shall use the descriptors `apparent' and `gravitational'
interchangeably, or even together---as in `the apparent (gravitational)
horizon'---throughout this paper.

Our principal goal here is to demonstrate why this horizon has become such an
essential ingredient in the interpretation of the cosmological data, and how
its measured value reveals the emergence of new physics beyond
the standard model of particle physics. Its relevance to our understanding of the
cosmos started to become quite evident when the optimization of free parameters in
the concordance model, $\Lambda$CDM \cite{Bennett:2003,Spergel:2003,Ade:2014},
revealed a very strange, unexpected coincidence---that the gravitational
radius today, $R_{\rm h}(t_0)$, equals $ct_0$ to within the measurement error
\cite{Melia:2003,Melia:2007,Melia:2012a}. As we shall see shortly, this equality
is not only surprising; it is actually highly improbable, and if we believe that
there is actual physics behind its emergence, the only way to account for it
is with a very special equation of state---the zero active mass condition,
$\rho+3p=0$ \cite{Melia:2016a,Melia:2017a}. It is specifically this constraint
that cannot accommodate a cosmological constant, replacing it instead with
a dark energy density that changes with time.

Before we begin our exploration of this unusual cosmological coincidence,
we should mention, as an aside, that there actually already exists another
confirming indication of the importance of $R_{\rm h}$, based on how it leads
to a resolution of the long-standing debate concerning the origin of cosmological
redshift, i.e., whether it constitutes a new form of time dilation, or merely
another instance of the better established kinematic/Doppler and gravitational
effects studied in other applications of general relativity \cite{Melia:2012b}.
The interpretation of cosmological redshift as due to an actual expansion of
space is problematic and almost certainly unphysical
\cite{Harrison:1995,Chodorowski:2007,Chodorowski:2011,Baryshev:2008,Bunn:2009,Cook:2009,Gron:2007}.
Instead, the introduction of the gravitational radius $R_{\rm h}$ proved that
it is simply a product of both the Doppler and gravitational redshifts in an
expanding cosmos \cite{Melia:2012b}.

\vspace{5mm}
\noindent{\large\bf 2  \  The $R_{\rm h}=ct$ Hypothesis}\\
To understand why the equality $R_{\rm h}(t_0)=ct_0$ is unexpected and highly
unlikely, let us begin by writing the spherically-symmetric FLRW metric in its
standard form, using comoving coordinates $(ct,r,\theta,\phi)$, where $t$ is the
cosmic time:
\begin{equation}
ds^2=c^2\,dt^2-a^2(t)[dr^2(1-kr^2)^{-1}+
r^2(d\theta^2+\sin^2\theta\,d\phi^2)]\;.
\end{equation}
Here, $a(t)$ is the universal expansion factor, so that the `proper' (or physical)
distance is $R\equiv a(t)r$, and the geometric constant $k$ is $+1$ for a closed
universe, $0$ for a flat, open universe, or $-1$ for an open universe.

Folding this metric through Einstein's field equations of general relativity,
one obtains the following equations of motion describing the cosmic expansion:
\begin{equation}
H^2\equiv\left({\dot a\over a}\right)^2={8\pi G\over 3c^2}\rho-{kc^2\over a^2}\;,
\end{equation}
known as the Friedmann equation \cite{Friedmann:1923}, and
\begin{equation}
{\ddot a\over a}=-{4\pi G\over 3c^2}(\rho+3p)\;,
\end{equation}
the so-called `acceleration' equation. $H$ is the Hubble constant, and an
overdot denotes a derivative with respect to $t$, while $\rho$ and $p$
represent, respectively, the total energy density and pressure, as noted
earlier.

The Friedmann equation is actually also derivable quite easily using the
Birkhoff theorem in the Newtonian limit \cite{Melia:2012c,Birkhoff:1923}, which
provides an intuitively satisfying meaning to the spatial curvature constant $k$.
One can show that it is proportional to the sum of local positive expansion
kinetic energy (related to $H^2$) and local negative potential energy
(represented by the term $8\pi G\rho/3c^2$). The observational data seem to indicate
that the Universe is spatially flat, meaning that $k=0$ \cite{Planck:2016}. Therefore,
the Big Bang apparently separated positive kinetic energy and negative potential energy
in equal (cancelling) portions, lending some support to the view that the Universe
may have begun its expansion as a quantum fluctuation in vacuum. Together with Equations~(1),
(2) and (4), this result also shows that
\begin{equation}
R_{\rm h}={c\over H}\;,
\end{equation}
which is readily recognizable as the Hubble radius. The measured value of the
Hubble constant today, $H_0\sim 70$ km s$^{-1}$ Mpc$^{-1}$, therefore suggests
a gravitational radius for the Universe of $R_{\rm h}\sim 14$ Glyr,
virtually indistinguishable from the distance $ct_0$ \cite{Planck:2016}.
But $R_{\rm h}(t_0)$ could have been anything---from as little as $1$ cm to as
large as $10^{100}$ times 14 Glyr---or more.

This range is enormous due to the flexibility of the solution to Equation~(4) with
$k=0$. The standard model of cosmology, known as $\Lambda$CDM, is based on the FLRW
metric with an empirically motivated choice of components in the energy density
$\rho$ of the cosmic fluid. It contains (visible and dark) matter ($\rho_{\rm m}$),
radiation ($\rho_{\rm r}$), and an unknown `dark' energy generally assumed to
be a cosmological constant $\Lambda$ ($\rho_\Lambda$), such that $\rho=
\rho_{\rm m}+\rho_{\rm r}+\rho_\Lambda$. Each of these constituents changes with
$a(t)$ in its own unique way, and conventional wisdom has it that radiation
dominated at early times, while matter and, more recently, dark energy have been
dominant since then. From basic physical principles we expect that
$\rho_{\rm m}\sim a^{-3}$ and $\rho_{\rm r}\sim a^{-4}$, while $\rho_\Lambda$
would be independent of $a$ if dark energy were truly a cosmological constant.
Therefore, defining the so-called critical density $\rho_{\rm c}\equiv 
3c^2H_0^2/8\pi G$, it is easy to see from Equation~(4) that
\begin{equation}
ct=R_{\rm h}(t_0)\int_0^a du\,\left(\Omega_{\rm m}u^{-1}+\Omega_{\rm r}u^{-2}
+\Omega_\Lambda u^2\right)^{-1/2}\;,
\end{equation}
where $\Omega_i\equiv \rho_i(t_0)/\rho_{\rm c}$ for each species ``i".
Using the conventional normalization $a(t_0)=1$, we thus find that
\begin{equation}
{R_{\rm h}(t_0)\over ct_0}=\left\{\int_0^1 du\,\left(\Omega_{\rm m}u^{-1}+
\Omega_{\rm r}u^{-2}+\Omega_\Lambda u^2\right)^{-1/2}\right\}^{-1}\;,
\end{equation}
which is heavily dependent on the density ratios $\Omega_{\rm m}$,
$\Omega_{\rm r}$ and $\Omega_\Lambda$. The problem is that these quantities
are entirely empirical, presumably set by initial conditions at the time of
the Big Bang, without any known theoretical underpinning. As far as we know,
in the context of $\Lambda$CDM they could have had any value.

But to be conservative, let us ignore the overall randomness in $H_0$,
$\Omega_{\rm m}$, $\Omega_{\rm r}$ and $\Omega_\Lambda$, and assume,
for whatever reason, that there actually does exist some unrecognized
theoretical basis for the values we have measured \cite{Planck:2016,Melia:2015a}, i.e.,
$H_0=67.4\pm0.5$ km s$^{-1}$ Mpc$^{-1}$, $\Omega_{\rm m}=0.315\pm 0.007$,
$\Omega_{\rm r}=(5.48\pm0.001)\times 10^{-5}$ and $\Omega_\Lambda=1.0-
\Omega_{\rm m}-\Omega_{\rm r}$. How would the ratio $R_{\rm h}(t)/ct$
as calculated from Equation~(8) with these {\it Planck} optimized
parameters change were we to measure it at times other than $t_0$?
The answer to this question is truly puzzling because, as shown in
Figure~1, even for this fixed set of parameters, $R_{\rm h}$ equals
$ct$ only once in the entire history of the Universe---and it must be
happening right now, at time $t_0$, just when we happen to be looking.
In other words, the fact that we infer the constraint $R_{\rm h}(t_0)=ct_0$
today is an astonishingly unlikely event. Actually, if the Universe's timeline is
infinite, as one might expect for an open FLRW model with $k=0$, the
probability of this happening at any particular time is essentially {\it zero}.

\begin{figure}
\begin{center}
\includegraphics[width=0.4\linewidth]{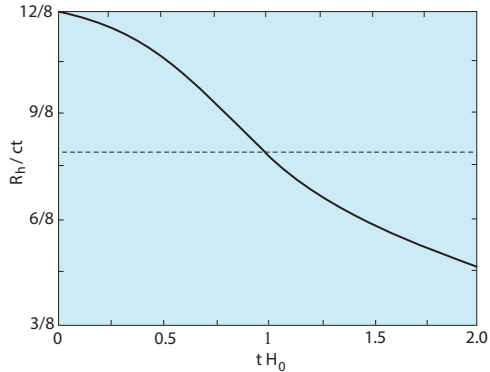}
\end{center}
\caption{The ratio $R_{\rm  h}(t)/ct$ as a function of $tH_0$, for the
291 {\it Planck} optimized parameters \cite{Planck:2016}, starting at the
292 Big Bang ($tH_0=0$) and extending to twice the current age of the
293 Universe. Since the spatial curvature constant is apparently zero
294 ($k=0$), the Universe is infinite, so the probability of us seeing
295 a ratio $R_{\rm  h}(t)/ct=1$ today ($tH_0=1$), just when we happen
296 to be looking, is essentially {\it zero}.}
\end{figure}

Equations~(5) and (8) are subject to all sorts of possible accelerations
and decelerations, yet the observations today are telling us that the
Universe's early phase of deceleration was exactly cancelled by a subsequent
acceleration, with a transition occurring roughly at the midpoint. Yet
this cancellation could not have occurred---and could not ever occur---at any
time other than $t_0$. For this reason, it is simply not sensible to accept
such an eventuality as a mere coincidence. There {\sl must} be some physics underlying
what we see.

Of course, there could be several possible explanations for this constraint,
but the simplest is simply that $R_{\rm h}$ is always equal to $ct$
\cite{Melia:2007,MeliaAbdelqader:2009,Melia:2012c}. Then any observer, no
matter when (s)he looks, would conclude that his apparent (gravitational) horizon
equals the distance light could have traveled since the Big Bang. In other
words, the probability of us seeing $R_{\rm h}(t_0)=ct_0$ today---or at any
other time---would then be one. As we shall see shortly, the observational
evidence in favour of the hypothesis $R_{\rm h}(t)=ct$ for all $t$ is now
quite compelling. Later in this paper, we shall highlight some of the most
striking supporting examples though, in reality, there have now been over
24 different kinds of test completed and published in the primary literature,
confirming this theory. But before we get to that discussion, let us first
consider a theoretical argument that supports the actual realization of
the zero active mass condition, for which the $R_{\rm h}(t)=ct$ constraint
is guaranteed for all $t$.

\begin{table*}
\tiny
  \caption{Observational Tests of the $R_{\rm h}=ct$ hypothesis and a comparison with $\Lambda$CDM}
  \centering
  \begin{tabular}{lll}
&& \\
    \hline
\hline
&& \\
Test or Observational Conflict& Outcome & Reference\\
&& \\
\hline
&& \\
Angular correlation function of the CMB& $R_{\rm h}=ct$ fits it very well; standard inflationary $\Lambda$CDM misses by $\sim 6 \sigma\quad$& \cite{MeliaCorredoira:2018}\\
Massive halo growth at $4\lesssim z\lesssim 10$ &Data consistent with $R_{\rm h}=ct$; $\Lambda$CDM misses by a factor $\sim 10^4$&
\cite{Steinhardt:2016,YennaMelia:2018}\\
Electroweak Horizon Problem & $R_{\rm h}=ct$ does not have it; $\Lambda$CDM currently has no solution&\cite{Melia:2018b}\\
Missing progenitors of high-$z$ quasars&In tension with $\Lambda$CDM, but consistent
with the timeline in $R_{\rm h}=ct$&\cite{FatuzzoMelia:2017}\\
Angular-diameter distance test with quasar cores& $R_{\rm h}=ct$ is favoured over $\Lambda$CDM with BIC
likelihood $81\%$ vs $19\%$&\cite{Melia:2018c,MeliaYenna:2018}\\
HII Hubble diagram&$R_{\rm h}=ct$ is favoured over $\Lambda$CDM with  BIC likelihood $93\%$
vs. $7\%$&\cite{Wei:2016,LeafMelia:2018a}\\
Alcock-Paczy\'nski test with the BAO scale & $R_{\rm h}=ct$ is favoured over $\Lambda$CDM at a $2.6\sigma$ confidence level&
\cite{MeliaCorredoira:2017} \\
FSRQ $\gamma$-ray luminosity function& $R_{\rm h}=ct$ is very strongly favoured over $\Lambda$CDM with $\Delta\gg10$ &\cite{Zeng:2016} \\
QSO Hubble diagram $+$ Alcock-Paczy\'nski& $R_{\rm h}=ct$ is about 4 times more likely than $\Lambda$CDM to be correct&
\cite{CorredoiraMelia:2016} \\
Constancy of the cluster gas mass fraction & $R_{\rm h}=ct$ is favoured over $\Lambda$CDM with BIC likelihood $95\%$ vs $5\%$&\cite{Melia:2016b} \\
Cosmic Chronometers&$R_{\rm h}=ct$ is favoured over $\Lambda$CDM with BIC likelihood $95\%$ vs $5\%$
&\cite{MeliaMaier:2013,MeliaMcClintock:2015a}\\
Cosmic age of old clusters&$\Lambda$CDM cannot accommodate high-$z$ clusters, but $R_{\rm h}=ct$ can&\cite{YuWang:2014}\\
High-$z$ quasars&The evolution timeline fits within $R_{\rm h}=ct$, but not $\Lambda$CDM&\cite{Melia:2013a,Melia:2018d,MeliaMcClintock:2015b} \\
The AGN Hubble diagram& $R_{\rm h}=ct$ is favoured over $\Lambda$CDM with BIC likelihood $96\%$ vs $4\%$&\cite{Melia:2015b} \\
Age vs. redshift of old passive galaxies& $R_{\rm h}=ct$ favoured over $\Lambda$CDM with BIC likelihood $80\%$ vs $20\%$&\cite{Wei:2015a} \\
Type Ic superluminous supernovae & $R_{\rm h}=ct$ is favoured over $\Lambda$CDM with BIC likelihood $80\%$ vs $20\%$&\cite{Wei:2015b} \\
The SNLS Type Ia SNe&$R_{\rm h}=ct$ is favoured over $\Lambda$CDM with BIC likelihood $90\%$ vs $10\%$&\cite{Wei:2015c} \\
Angular size of galaxy clusters&$R_{\rm h}=ct$ is favoured over $\Lambda$CDM with BIC likelihood $86\%$ vs $14\%$&\cite{Wei:2015d} \\
Strong gravitational lensing galaxies&Both models fit the data very well due to the bulge-halo `conspiracy'& \cite{Melia:2015c,LeafMelia:2018b}\\
Time delay lenses&$R_{\rm h}=ct$ is favoured over $\Lambda$CDM with BIC likelihood $80\%$ vs $20\%$&\cite{Wei:2014a} \\
High-$z$ galaxies&The evolution timeline fits within $R_{\rm h}=ct$, but not $\Lambda$CDM&\cite{Melia:2014a} \\
GRBs $+$ star formation rate&$R_{\rm h}=ct$ is favoured over $\Lambda$CDM with AIC likelihood $70\%$ vs $30\%$&\cite{Wei:2014b} \\
High-$z$ quasar Hubble diagram&$R_{\rm h}=ct$ is favoured over $\Lambda$CDM with BIC likelihood $85\%$ vs $15\%$&\cite{Melia:2014b} \\
GRB Hubble diagram&$R_{\rm h}=ct$ is favoured over $\Lambda$CDM with BIC likelihood $96\%$ vs $4\%$&\cite{Wei:2013} \\
&& \\
\hline\hline
  \end{tabular}
\end{table*}

\vspace{5mm}
\noindent{\large\bf 3  \  The Lapse function in FLRW Cosmologies}\\
The $R_{\rm h}=ct$ hypothesis has recently been validated by a
theoretical study of the lapse function $g_{tt}$ in the FLRW metric.
This is the coefficient multiplying the $c^2\,dt^2$ term in Equation~(3).
As is well known, the FLRW metric adopts a high degree of
symmetry, including homogeneity and isotropy, making it a special member
of the class of spherically-symmetric spacetimes describing gravitational
collapse or expansion \cite{Oppenheimer:1939,McVittie:1964,Misner:1964,Thompson:1967}.
Unlike the other metrics in this category, however, the lapse function
$g_{tt}$ in FLRW is always set equal to one---regardless of which equation
of state is assumed for the stress-energy tensor $T^{\mu\nu}$.

While setting $g_{tt}=1$ is consistent with the Cosmological principle
(i.e., the assumption of homogeneity and isotropy), however, there is actually
no justification for supposing that the lapse function should always be
coordinate-independent for any arbitrary stress-energy tensor $T^{\mu\nu}$.
For a metric with $g_{tt}=1$, the observer sees no time dilation,
regardless of whether or not the cosmic expansion is accelerating. This
issue should be problematic at a fundamental level in GR,
but is often ignored in the literature because $g_{tt}$ in FLRW
can, at most, be a function only of time, not space, in order for it to be
consistent with isotropy. Therefore, conventional wisdom has it that one
can, if necessary, change the gauge $dt\rightarrow dt^\prime\equiv\sqrt{g_{tt}}\,dt$
to restore a coordinate-independent lapse function in FLRW, while retaining
one's preferred expansion factor $a(t)$.

But this complacency has been challenged recently with an application of the Local Flatness
Theorem (see, e.g., refs.~\cite{Weinberg:1972}) to examine whether
`creating' such a frame of reference is in fact consistent with the GR mandated
existence of a local free-falling frame against which one can always measure the
spacetime curvature and concomitant time dilation in the given metric
\cite{Melia:2019}. This is a somewhat subtle point, but fundamentally rooted
in the equivalence principle. The Local Flatness Theorem shows that the conflict
between a unitary value of the lapse function and an arbitrary choice of $a(t)$
cannot be removed with a simple change of gauge which, in relativity, is actually
a transformation of the coordinates, i.e., a change in reference frames
\cite{Melia:2016a}.

The Local Flatness Theorem compels \cite{Melia:2019} the FLRW lapse function
to satisfy the following constraint:
\begin{equation}
\int^{ct} g_{tt}(t^\prime)\,d(ct^\prime) = cg_{tt}(t){a\over \dot{a}}\;.
\end{equation}
Therefore, the choice of lapse function $g_{tt}(t)=1$ is consistent only
with the expansion factor
\begin{equation}
a(t) \equiv \left({t\over t_0}\right)\;,
\end{equation}
normalized in conventional form for a spatially flat metric with
$a(t_0)=1$. (There is, of course, also the trivial solution $a=$ constant,
for which the spacetime curvature is identically zero, and the coordinates
of the FLRW metric in Equation~3 are then those in Minkowski space.)

A quick inspection of the acceleration Equation~(5) therefore shows that there is only
one equation of state consistent with the choice of lapse function $g_{tt}=1$
(other than the inconsequential $a=$ constant case, for which $\rho=p=0$), and
this is the so-called zero active mass condition, $\rho+3p=0$, which produces
a constant universal expansion rate. But this is not the empty Milne universe,
in which $\rho=p=0$ and the linear expansion is driven by spatial curvature
with $k=-1$. It is trivial to see from Equation~(4) that a flat FLRW Universe
must always have a Hubble constant $H(t)=1/t$ and an apparent (gravitational)
radius $R_{\rm h}(t)=ct$.

What appeared to be an unlikely coincidence in Figure~1 turns out to
be well justified theoretically. One may wonder, then, how the standard model
of cosmology, which has always ignored the role played by $R_{\rm h}$,
successfully accounts for many of the observations. As we shall see below,
$\Lambda$CDM has served us well with gross interpretations of the data, but
not so well once the measurement precision started to improve. Today, the
tension between its predictions and the observations is too large to ignore
and the disparity between standard theory and experiment in several areas
is quite glaring. In subsequent sections of this paper, we shall also
highlight the contrasting success of the $R_{\rm h}=ct$ hypothesis in fixing
(or removing) all of these emerging problems.

\vspace{5mm}
\noindent{\large\bf 4  \  Observational Tests of the $R_{\rm h}=ct$ Hypothesis}\\
An initial assessment of the preceeding discussion might seem to be at odds
with the standard model's success in accounting for a broad range of cosmological
data. After all, $\Lambda$CDM contains dark energy in the form of a cosmological
constant, $\Lambda$, with pressure $p_\Lambda=-\rho_\Lambda$ and a total equation
of state based on individual variations of the energy density for each species.
Therefore, $p=p_{\rm m}+p_{\rm r}+p_\Lambda$ in this model is unavoidably an evolving
function of $\rho=\rho_{\rm m}+\rho_{\rm r}+\rho_\Lambda$.

It is thus crucial, not only to gauge how the $R_{\rm h}=ct$ equality fares on
average over a Hubble time but, to simply impose the zero active mass condition
on $\Lambda$CDM for all $t$ and examine whether it mitigates or eliminates the
tension growing between theory and observation. This effort has been underway
for over a decade now, including over 24 different kinds of measurement, at low
and high redshifts, using integrated times and distances, and also the
differential rate of expansion, $H(z)$. A summary of the completed tests is
provided in Table~1, along with references to the literature where the results
have been published. It is quite remarkable that---perhaps contrary to
expectation---the equation of state $\rho+3p=0$ reduces or eliminates any
disparity between the data and model predictions in every case studied thus far.

The outcomes shown in this table are based on model selection
tools designed to `choose' which model (or models) are preferred by the data.
The Akaike Information Criterion (AIC \cite{Liddle:2007}), the Kullback Information
Criterion (KIC \cite{Cavanaugh:2004}), and the Bayes Information Criterion (BIC
\cite{Schwarz:1978}) are used often for this purpose. In terms of the
maximum value of the likelihood function $\mathcal{L}$ and the number $n$
of free parameters in the model, the difference $\Delta_{\rm IC}=
{\rm IC}_2-{\rm IC}_1$ determines the extent to which model $\mathcal{M}_1$
is preferred over model $\mathcal{M}_2$, with ${\rm IC}_\alpha=
-2\ln\mathcal{L}_\alpha+2n_\alpha$, and IC is either AIC, KIC, or BIC, as
the case may be \cite{MeliaMaier:2013}. The outcome $\Delta_{\rm IC}$ is considered
to represent `positive' evidence that model 1 is preferred over model 2 when
$\Delta_{\rm IC}>2$; `moderate' when $2<\Delta_{\rm IC}<6$; and very strong when
$\Delta_{\rm IC}>10$. One may also use $\Delta_{\rm IC}$ to estimate the relative
probability (or percentage likelihood) that model 1 is statistically preferred over
model 2, according to the formulation
\begin{equation}
P(\mathcal{M}_1) = {1\over 1+\exp{(-\Delta_{\rm IC}/2)}}\;,
\end{equation}
with $P(\mathcal{M}_2)=1-P(\mathcal{M}_1)$, in a head-to-head comparison
between two models.

\begin{figure}
\begin{center}
\includegraphics[width=0.6\linewidth]{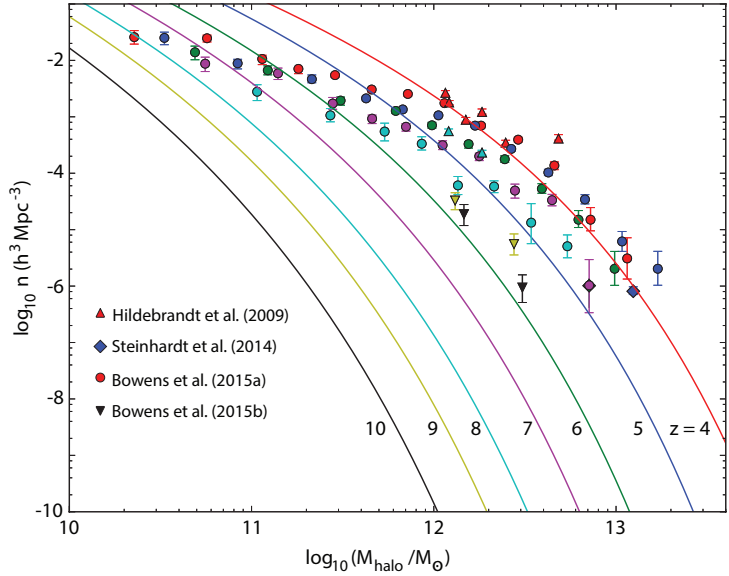}
\end{center}
\caption{Halo mass function inferred from galaxy surveys, as
a function of mass and redshift: $z=4$ (red), $5$ (blue), $6$ (green),
$7$ (magenta), $8$ (cyan), $9$ (yellow), and $10$ (black). Solid curves
represent the theoretical halo number density predicted by $\Lambda$CDM in
this same redshift range, based on the estimates of ref.~\cite{Sheth:2001} and
calculated with the HMFCalc code of ref.~\cite{Murray:2013}. (Adapted from
ref.~\cite{Steinhardt:2016})}
\end{figure}

Table~1 shows that all of the comparative tests concluded thus far consistently
favour $R_{\rm h}=ct$ over basic $\Lambda$CDM without the zero active mass equation
of state. The preference is sometimes moderate, and often strong. For illustration
we here highlight two of the most prominent examples where continued support for
basic, inflationary $\Lambda$CDM over $R_{\rm h}=ct$ appears to be untenable.
These are (1) the inferred halo distribution as a function of mass and redshift
in the redshift range $4\lesssim z\lesssim 10$, which disagrees with structure
formation in $\Lambda$CDM by as much as four orders of magnitude; and (2) the
angular correlation function of the CMB, which shows strong evidence of a minimum
cutoff in the power spectrum, conflicting with the basic inflationary paradigm.

\begin{figure}
\begin{center}
\includegraphics[width=0.7\linewidth]{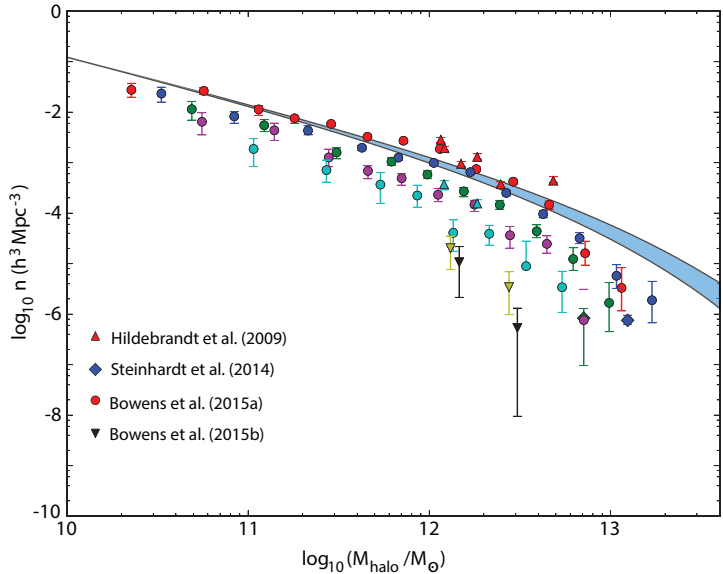}
\end{center}
\caption{Same as figure~2, except that the data have been recalibrated
for the $R_{\rm h}=ct$ universe using the ratio of differential comoving volumes
\cite{YennaMelia:2018}.  The solid curves represent the halo mass function for
$R_{\rm h}=ct$, using a normalization based on the optimized fit to recently
published redshift space distortion measurements of the cosmological growth rate
\cite{Melia:2017b}. The halo mass function in $R_{\rm h}=ct$ is essentially
independent of redshift in the range $4\lesssim z\lesssim 10$. (Adapted from
ref.~\cite{YennaMelia:2018})}
\end{figure}

The halo mass data are shown in figure~2, together with seven theoretical
curves calculated for $4\lesssim z\lesssim 10$, using the halo mass function
estimates of ref.~\cite{Sheth:2001}, and the HMFCalc code developed in
ref.~\cite{Murray:2013}. The $\Lambda$CDM parameters are assumed to have their
{\it Planck} values (see \S~II above and ref.~\cite{Planck:2016}).
The disparity between the predicted and inferred distributions is quite
pronounced. The surprisingly early
appearance of massive galaxies significantly challenges the standard model,
and the halo mass function at $z\gtrsim 4$ is grossly inconsistent with the
predictions of $\Lambda$CDM, a situation termed ``The Impossibly Early Galaxy
Problem" by workers in the field \cite{Steinhardt:2016}. The most significant
tension is caused by the lack of anticipated strong evolution in redshift and
the absence of a steepening of the distribution with increasing mass.

The unexpected nature of the halo mass function has also been characterized
as a problem with the theory of galaxy formation at high $z$, but this
outcome is not new. The observed high-redshift quasars and galaxies would
have formed much too quickly compared to the timeline in the standard model
\cite{Melia:2013a,Melia:2014a}. To account for the emergence of $\sim 10^9\;
M_\odot$ black holes earlier than $z\sim 6-7$ in $\Lambda$CDM, one must
assume that black holes started growing from anomalously large seeds
($M> 10^5\;M_\odot$), or evolved at super-Eddington rates, neither of
which has ever been justified on astrophysical grounds. By comparison,
the timeline in $R_{\rm h}=ct$ would have allowed these high-redshift
objects to form following well understood astrophysical
principles \cite{Melia:2013a,Melia:2014a,Melia:2018d}.

Insofar as the halo masses are concerned, linear perturbation
theory in $R_{\rm h}=ct$ predicts the distribution shown in figure~3
\cite{YennaMelia:2018}.  The dependence on mass and redshift is
far more compatible with the data than the corresponding prediction
by $\Lambda$CDM. There is still work to be done, however, because
the anticipated distribution still deviates from the observations by
some unidentified systematic effect at the highest masses.

\begin{figure}
\begin{center}
\includegraphics[width=0.5\linewidth]{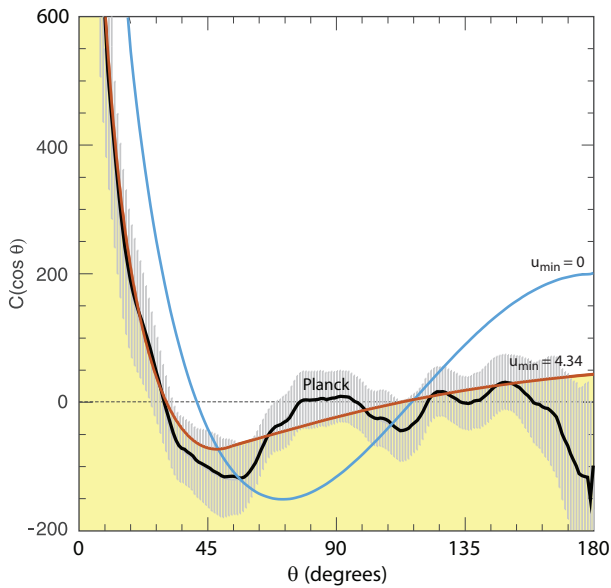}
\end{center}
\caption{Angular correlation function measured with {\it Planck}
(dark solid curve) \cite{Planck:2014}, including $1\sigma$ errors
(grey), compared with (blue) the conventional inflationary $\Lambda$CDM
prediction, and (red) truncated inflation, or a non-inflationary
cosmology \cite{MeliaCorredoira:2018}. The parameter $u_{\rm min}$ is
proportional to the minimum wavenumber in the power spectrum. A value
of $0$, predicted by inflationary $\Lambda$CDM, is ruled out at a
level of significance greater than $6\sigma$. In $R_{\rm h}=ct$, the
measured value of $u_{\rm min}$ roughly corresponds to the Planck
scale, where the quantum fluctuations enter the classical Universe
\cite{Melia:2017c}.}
\end{figure}

A comparably serious problem has emerged in the angular correlation function
of the CMB, starting with the {\it Wilkinson} Microwave Anisotropy Probe (WMAP)
\cite{Bennett:2003,Spergel:2003} and resoundingly confirmed by {\it Planck}
\cite{Planck:2014}. This is but one of several anomalies seen on very large scales,
though the lack of any significant correlation seen at angles $\gtrsim 60^\circ$
is easily the most glaring one. The fact that it is inconsistent with the
basic inflationary concept has understandably initiated vigorous debate
about whether it is real, or due to some unknown observational systematic
effect. It could, e.g., be due to an improper foreground subtraction
\cite{Bennett:2013}, but after three independent missions have all
confirmed this problem (COBE preceded WMAP and {\it Planck}), it is difficult
to accept that subtleties in the foreground subtraction have yet to be
resolved following three decades of observation. The lack of large-angle
correlations could also be due to cosmic variance \cite{Bennett:2013,Copi:2015},
but the probability of this actually happening is typically $\lesssim 0.24\%$,
showing an inconsistency with the inflationary paradigm at better than
$3\sigma$ \cite{Kim:2011,Melia:2014c,Gruppuso:2016}.

This disparity is quite serious because---whereas anisotropies measured at
angles $\lesssim 1^\circ$ are mostly due to local astrophysical processes
(such as acoustic wave propagation)---the large-scale fluctuations reflect the
physics of dynamical expansion, i.e., the cosmological model itself.
To study this problem in greater detail, a re-analysis of the large-scale
anisotropies using a recent {\it Planck} release \cite{Planck:2014} therefore
took a different approach \cite{MeliaCorredoira:2018}. Instead of following
convention established by the inflationary picture, this assumed that quantum
fluctuations generated in the early Universe had a well-defined power spectrum
$P(k)$ with a wavenumber cutoff $k_{\rm min}\not=0$. Basic inflationary models
do not have such a physical scale, so if the CMB data rule out a zero $k_{\rm min}$,
this would be compelling evidence against the inflationary solution to the horizon problem
and, by association, the early phase of decelerated expansion predicted by $\Lambda$CDM.

Figure~4 shows the CMB angular correlation function measured in this recent
analysis (solid black), together with the $\Lambda$CDM prediction (blue) and
the optimized fit (red) using a non-zero $k_{\rm min}$. The parameter
$u_{\rm min}$ is proportional to $k_{\rm min}$ and is defined as follows:
$u_{\rm min}\equiv k_{\rm min}\,c\Delta\tau_{\rm dec}$, where $c\Delta\tau_{\rm dec}$
is the comoving radius of the last scattering surface (at `decoupling') in terms of
the conformal time difference between $t_0$ and $t_{\rm dec}$. The lack
of adequate confirmation by the data of the standard model prediction is
difficult to hide. The best-fit (red) curve corresponds to an optimized
value $u_{\rm min}=4.34\pm 0.50$, implying a maximum fluctuation
size $\theta_{\rm max}\approx 83^\circ$ in the plane of the sky. In
$\Lambda$CDM, decoupling is thought to have occurred at $z_{\rm dec}=1080$,
so this measurement of $u_{\rm min}$ corresponds to a maximum fluctuation
wavelength $\lambda_{\rm max}\sim 20$ Mpc at that redshift.

There are several reasons why the red curve is strongly preferred over the
standard model's prediction, including the fact that it not only accounts
for the lack of large-angle fluctuations, but actually fits the measurements
to within $1\sigma$ at {\it all} angles; and it correctly accounts for the
amplitude of the minimum of the angular correlation function and the angle
at which it occurs. Most importantly, these results show that a non-zero
cutoff in $P(k)$ is favoured over $k_{\rm min}=0$ at a confidence level
exceeding $\sim 8\sigma$.

But while a $k>k_{\rm min}$ constraint on $P(k)$ is inconsistent with slow-roll
inflationary cosmology, it is actually expected with the $R_{\rm h}=ct$ hypothesis
\cite{Melia:2007,Melia:2012c,Melia:2017c}. A viable scenario with this picture
has the fluctuations emerging at or near the Planck scale, which equals the
apparent (gravitational) horizon at the Planck time. The non-zero value of
$k_{\rm min}$ therefore corresponds to the size of this horizon at $t_{\rm dec}$,
since only fluctuations with a size $\lesssim \lambda_{\rm max}\sim 2\pi R_{\rm h}$
would have grown continuously towards the last scattering surface. In the
next section, we shall discuss the consequences of this work on the viability
of the inflationary paradigm in more detail, together with another `horizon'
problem now emerging, based on the vacuum expectation value of the Higgs
field.

\vspace{5mm}
\noindent{\large\bf 5  \  New Physics}\\
As we have already hinted earlier, and will demonstrate
more formally in \S~V.B below, a universe with $R_{\rm h}=ct$ cannot
be driven by a blended equation of state for matter, radiation and a
cosmological constant. This exclusion has several significant consequences,
including (1) that it implies a need to extend particle
physics beyond the standard model, and (2) that it provides a simple, elegant
solution to the so-called horizon problem, arising from the uniformity of
{\sl both} the CMB temperature across the sky and the expectation value of
the Higgs field. We begin our discussion with the latter, which has plagued
theoretical cosmology for over half a century.

\begin{figure}
\begin{center}
\includegraphics[width=0.4\linewidth]{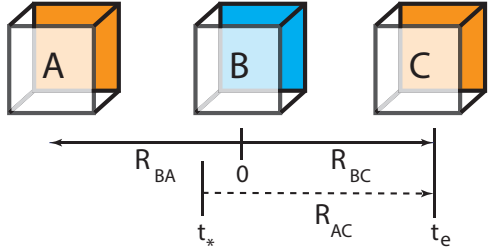}
\end{center}
\caption{Observer B receives light signals from two
regions (A and C) in the CMB. Light was emitted by these
sources at cosmic time $t_e$, a proper distance $R_{BA}(t_e)=R_{BC}(t_e)$
from the observer. Patch (A) emitted a light signal at the earlier time $t_*$
that reached (C), a proper distance $R_{AC}(t_e)$ away, at time $t_e$. (Adapted from ref.~\cite{Melia:2013b})}
\end{figure}

\vspace{5mm}
\noindent{\bf 5.1  \ The Temperature and Electroweak Horizon Problems}\\
The CMB temperature horizon problem is a major shortcoming of $\Lambda$CDM
because---without some anomalously accelerated expansion at early
times---the Universe would have required highly improbable, specialized initial
conditions. The microwave background radiation has the same temperature
everywhere, except for tiny, random fluctuations of one part in 100,000.
We infer that opposite sides of the sky (patches A and C in figure~5) lie
beyond each other's horizon (which is actually not completely correct, but
this is not the principal difficulty), yet their temperatures today are
identical.

\begin{figure}
\begin{center}
\includegraphics[width=0.6\linewidth]{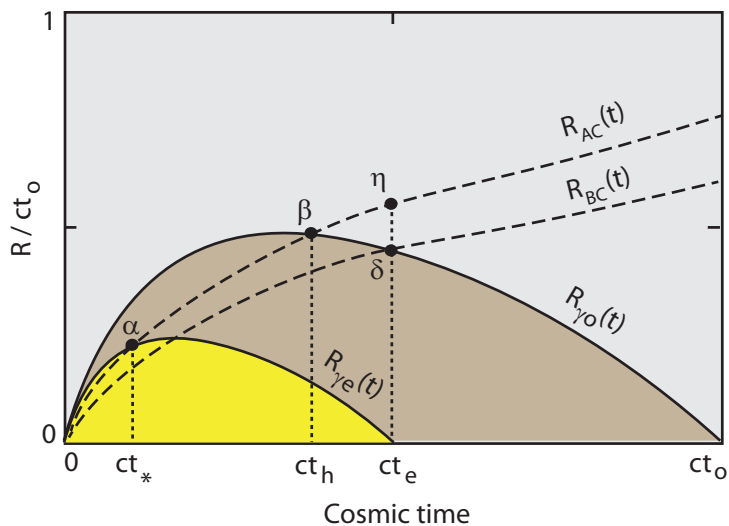}
\end{center}
\caption{Proper distance $R_{\gamma 0}$ ($R_{\gamma\rm e}$) of light approaching
an observer in patch (C) at time $t_0$ ($t_{\rm e}$) as a function of $t$,
and the corresponding proper distances ($R_{\rm AC}$ and $R_{\rm BC}$) of (A) and (B)
relative to (C), all calculated in the $\Lambda$CDM cosmology. (Adapted from
ref.~\cite{Melia:2013b})}
\end{figure}

The temperature horizon problem in $\Lambda$CDM is best understood by
studying the diagram in figure~6. For an observer at (C), the curves
labeled $R_{\gamma 0}$ and $R_{\gamma\rm e}$ show the null geodesics
reaching him/her at times $t_0$ and $t_{\rm e}$, respectively. Correspondingly,
the curves labeled $R_{\rm AC}$ and $R_{\rm BC}$ give the proper distances
between (A) and (C), and between (B) and (C), as functions of time. So
light emitted in patch (B) at spacetime point $\delta$ would reach $C$
today (at time $t_0$), while light emitted in patch (A) at $\alpha$
would have reached (C) at time $t_e$. Light emitted by one of them (say
patch A), at $t_*<t_{\rm e}$, could have reached (C) prior to the emission
of the CMB at $t_{\rm e}$, but in $\Lambda$CDM there is no value of $t_*\ge 0$
for which $R_{\rm AC}(t_{\rm e})=2R_{\rm BC}(t_{\rm e})$, with an $R_{\rm BC}(t_0)$
bigger than the proper size of our visible Universe \cite{Melia:2013b}.
Therefore, an observer at (B) would not able to see a uniform temperature today.

\begin{figure}
\begin{center}
\includegraphics[width=0.6\linewidth]{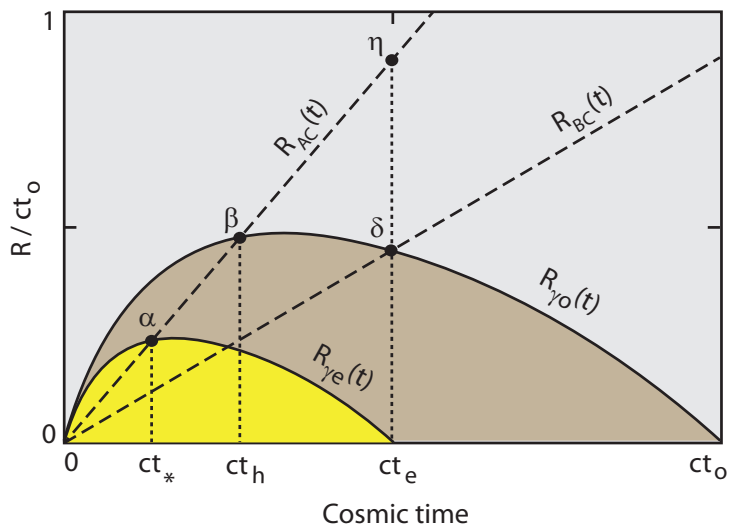}
\end{center}
\caption{Same as figure~6, except now for a universe with
$R_{\rm h}=ct$, i.e., with the zero active mass equation of state.
Note that, unlike the situation in figure~6, this time $R_{\rm AC}(t_e)=
2R_{\rm BC}(t_e)$. (Adapted from ref.~\cite{Melia:2013b})}
\end{figure}

The root of this problem is simply that the early Universe in $\Lambda$CDM
decelerated following the Big Bang, preventing patches (A) and (C) from
receeding sufficiently from (B) prior to producing the CMB at $t_{\rm e}$.
But this problem is completely absent in $R_{\rm h}=ct$ because $R_{AC}$
was {\sl always} equal to $2R_{BC}$ (see figure~7). So no matter the time
$t_{\rm e}$ at which decoupling took place, one can always find a $t_*>0$
at which the emission of a signal by either patch could have reached and
equilibrated the physical conditions in the other. In other words,
$R_{\rm h}=ct$ never had a horizon problem to begin with.

This great simplification and avoidance of a fundamental inconsistency
in the expansion dynamics is a major strength of the $R_{\rm h}=ct$
hypothesis, more so, because $\Lambda$CDM without the zero active mass
equation of state is facing a daunting task to overcome its horizon
problems. The most likely solution comes from particle physics, which
suggests that several phase transitions might have occurred,
including an inflationary event
from the separation of the strong and electroweak forces in grand unified
theories (GUTs) \cite{Guth:1981,Linde:1982}. The accelerated expansion
at $t\sim 10^{-36}-10^{-33}$ seconds might have greatly enlarged the
proper distances $R_{AC}(t_{\rm e})$ and $R_{BC}(t_{\rm e})$ in figure~6
beyond our causal horizon, allowing us to see a uniform CMB temperature
everywhere. Even after three decades of development, however, we still
do not have a complete, self-consistent theory of inflation
\cite{Ijjas:2013,Ijjas:2014}. We do not know the inflaton potential, nor
do we have the trans-Plancking physics necessary to describe how the
inflaton field could have emerged into the classical Universe from its
remote early beginning inaccessible to modern quantum mechanics.

Now that the Higgs particle has been discovered, the situation for
$\Lambda$CDM is much worse, because the Universe probably passed through
another (electroweak) phase transition at $T=159.5\pm1.5$ GeV---about
$\sim 10^{-11}$ seconds following the Big Bang. This `turning on' of
the Higgs field gave mass to the fermions and separated the weak and
electric forces. The problem is that the vacuum expectation value of the
Higgs field is expected to be uniform only within a region that was
causally connected at the time of the phase transition. In $\Lambda$CDM,
that region would have expanded to about $40$ Mpc today \cite{Melia:2018b}.
The proper size of our visible Universe, however, is approximately
$2,212$ Mpc, roughly $50$ times larger, yet no evidence of variable
particle properties, such as lepton mass, has ever been seen.
Unlike the GUT transition that might have produced inflation to fix
the CMB temperature horizon problem, there is no known solution to
this equally troubling Higgs horizon inconsistency. The Higgs horizon
problem is actually worse than the CMB temperature inconsistency because
Higgs physics is now understood better than GUTs.

\vspace{5mm}
\noindent{\bf 5.2  \ Dynamical Dark Energy}\\
With the $R_{\rm h}=ct$ hypothesis, the cosmic fluid must maintain
a fixed (total) equation-of-state $w\equiv p/\rho=-1/3$. Thus, given
the dependence of $\rho_{\rm m}$ and $\rho_{\rm r}$ on the expansion
factor $a(t)$ (see \S~II), it is not possible to do this with a
cosmological constant, whose pressure is fixed at $p_\Lambda=-\rho_\Lambda$.
As we shall see, the dark energy density must evolve along with the
other particle species, so it must be dynamic---perhaps a new particle
(or particles) in extensions to the standard model. In this section, we
estimate how $\rho_{\rm de}$ and $p_{\rm de}$ must have changed over a
Hubble time in order to comply with the zero active mass condition.

Putting $\rho=\rho_{\rm r}+\rho_{\rm m}+\rho_{\rm de}$, and
$p=-\rho/3=w_{\rm de}\rho_{\rm de}+\rho_{\rm r}/3$, we immediately see that
\begin{equation}
\rho_{\rm r}=-3w_{\rm de}\rho_{\rm de}-\rho\;,
\end{equation}
under the assumption that $p_{\rm r}=+\rho_{\rm r}/3$ and $p_{\rm m}\approx 0$.
From the Friedmann Equation~(4) with spatial flatness ($k=0$), we also
recognize that, throughout cosmic history, the overall density has evolved
according to
\begin{equation}
\rho(t)=\rho_{\rm c}\,a(t)^{-2}\;,
\end{equation}
where $\rho_{\rm c}$ is the previously defined critical density.

\begin{figure}
\begin{center}
\includegraphics[width=0.6\linewidth]{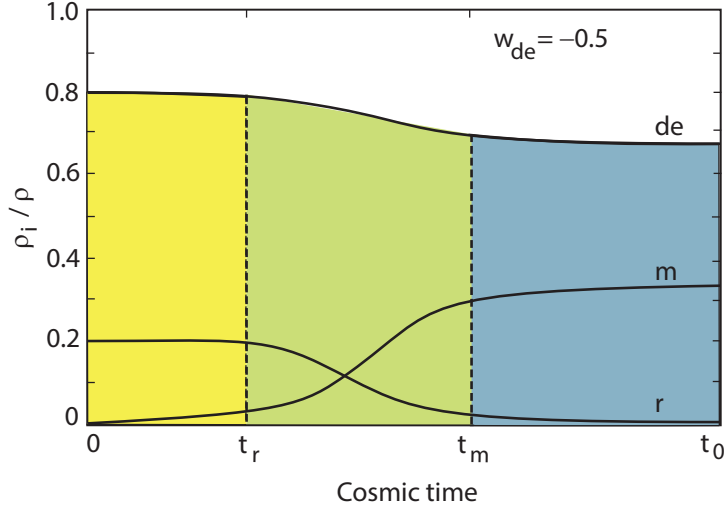}
\end{center}
\caption{Schematic diagram of the possible evolution in time of various
energy densities $\rho_i$ in $R_{\rm h}=ct$: dark energy (de),
radiation (r) and matter (m). The empirical evidence today suggests
that $w_{\rm de}=-0.5$, so $\rho_{\rm r}/\rho=0.2$ and $\rho_{\rm de}/\rho=0.8$
at $z\gg 1$, while $\rho_{\rm m}/\rho=1/3$ and $\rho_{\rm de}/
\rho=2/3$ for $z\sim 0$. The temperature and overall density increase
back to the Big Bang, so radiation dominates over matter early on (i.e.,
in the region $t< t_{\rm r}$), while matter dominates over
radiation at late times (i.e., for $t>t_{\rm m}$).}
\end{figure}

Equation~(12) is valid at any epoch. At low redshifts, however,
the empirical evidence tells us that the CMB temperature ($T_0\approx 2.728$ K)
translates into a negligibly small normalized radiation energy density,
$\Omega_{\rm r}\sim 5\times 10^{-5}$, compared to matter and dark energy.
It is easy to see from the definitions of $\rho$ and $p$ that $w_{\rm de}$
must therefore be $\sim -1/2$ in order to reflect the partitioning of
$\rho_{\rm m}$ and $\rho_{\rm de}$ that we observe in the
local Universe. For convenience, we shall simply put $w_{\rm de}=-1/2$.
In that case,
\begin{equation}
\Omega_{\rm de}\approx -{1\over 3w_{\rm de}}\approx {2\over 3}\;,
\end{equation}
and
\begin{equation}
\Omega_{\rm m}\approx {1+3w_{\rm de}\over 3w_{\rm de}}\approx {1\over 3}
\end{equation}
in which, of course, $\Omega_{\rm m}$ is the sum of both baryonic
and dark matter, i.e., $\Omega_{\rm m}=\Omega_{\rm b}+\Omega_{\rm d}$
\cite{MeliaFatuzzo:2016}.

As noted earlier, we expect that radiation becomes dominant over
matter at large redshifts. Radiation on its own, however, cannot
produce a zero active mass equation of state. Thus, to comply with
the $R_{\rm h}=ct$ constraint, dark energy must be present along
with radiation in the early Universe. We thus put $\rho\approx 
\rho_{\rm r}+\rho_{\rm de}$. In that case, we see that
\begin{equation}
\rho_{\rm de}\approx {2\over 1-3w_{\rm de}}\rho_{\rm c}(1+z)^2\quad (z\gg 1)\;,
\end{equation}
and
\begin{equation}
\rho_{\rm r}\approx {3w_{\rm de}+1\over 3w_{\rm de}-1}\rho_{\rm c}(1+z)^2\quad (z\gg 1)\;,
\end{equation}
and if we also assume that $w_{\rm de}=-1/2$ remains constant throughout
cosmic evolution, then $\rho_{\rm de}=0.8\rho$ and $\rho_{\rm r}= 0.2\rho$
at high redshifts. These trends are shown schematically in figure~8.

This simple argument suggests that the zero active mass equation of
state $\rho+3p=0$ requires a gradual transition of the equilibrium
partitioning of the various constituents. We infer that, as a fraction
of the total energy density, dark energy decreases fractionally from
$\rho_{\rm de}/\rho=0.8$ when $z\gg 1$, to $\rho_{\rm de}/\rho=2/3$
at the present. During this evolution, radiation gradually dilutes,
from $\rho_{\rm r}/\rho= 0.2$ in the early Universe, to an insignificant
fraction at the present time, and the fractional representation of
matter increases from an insignificant amount at $z\gg 1$ to
$\rho_{\rm m}/\rho= 1/3$ at $z\rightarrow 0$.

These couplings and interactions provide us with some clues into the
nature of dark energy in $R_{\rm h}=ct$. As we see in figure~8, the radiation
energy density could not have diluted according to $(1+z)^4$ at high
redshifts, so it may have coupled to dark energy. Its designation as
`dark' would then depend on redshift, since it may have coupled to
the electromagnetic (or even the electroweak) field at very high
densities and temperatures. In the transition
region $t_{\rm r}\lesssim t\lesssim t_{\rm m}$, the gradual decrease
in the ratio $\rho_{\rm de}/\rho$ and increase in $\rho_{\rm m}/\rho$
may signal a partial decay of dark energy into standard model particles.
If the dark energy fraction is plateauing now, perhaps this is also an
indication that there were multiple dark energy particles, one or more
of which completely decayed prior to cosmic time $t_{\rm m}$, while the
rest have remained more stable.

\vspace{5mm}
\noindent{\large\bf 6  \  Conclusions and Future Prospects}\\ 
In this paper, we have argued for the paramount importance of the
apparent (gravitational) horizon in cosmology, whose direct influence
on our inerpretation of the data suggests a zero active mass equation
of state $\rho+3p=0$ in the cosmic fluid. This conclusion is based, at
least initially, on the most likely explanation for the highly improbable
coincidence of us seeing $R_{\rm h}(t_0)=ct_0$ today. The solution would
simply be that this equality is actually maintained at all times. More
recent theoretical work has shown that this outcome may ultimately have a
fundamental basis in general relativity, which limits the applicability
of the FLRW metric to stress-energy tensors that allow only a
non-accelerated expansion of the Universe. As we now understand,
a cosmos with a constant expansion rate necessarily satisfies the
$R_{\rm h}=ct$ constraint on its apparent (gravitational) horizon.

But how should one understand the dynamics of dark energy, whose
own equation of state is just right to ensure the zero active mass
condition overall? Given the panoply of clues from nature, it seems
reasonable to assume that the single most important feature of the
Big Bang was the magnitude of the separated (positive) expansion
and (negative) potential energies. This initial condition directly
affects the Hubble constant, and all observational signatures that
stem from it, including the angular-diameter and luminosity distances,
and age versus redshift relation, among many others.

\begin{figure}
\begin{center}
\includegraphics[width=0.5\linewidth]{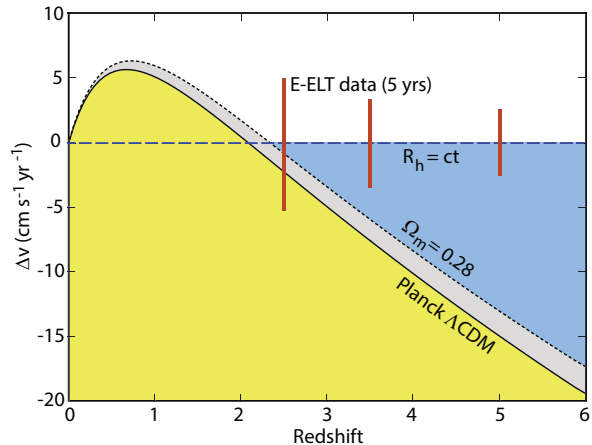}
\end{center}
\caption{Predicted velocity shift $\Delta v$ underlying the redshift drift for
three cosmological models \cite{Melia:2016c}: (solid black) {\it Planck} $\Lambda$CDM
($k=0$, $\Omega_m=0.315$ $H_0=67.4$ km s$^{-1}$ Mpc$^{-1}$) \cite{Planck:2016};
(black dashed) a slight variation of $\Lambda$CDM with $\Omega_m=0.28$; and
(blue long dash) the $R_{\rm h}=ct$ hypothesis, for which $\Delta v=dz/dt_0=0$
at all redshifts. The red bars indicate the expected $1\sigma$ errors at
$z=2.5$, $3.5$, and $5.0$ with the ELT-HIRES after 5 years of monitoring.
(Adapted from ref.~\cite{Martins:2016})}
\end{figure}

If we further maintain that the Universe ought to be homogeneous and
isotropic, and simultaneously adopt Weyl's postulate (which essentially
says that no two worldlines may ever cross), general relativity leaves
us with only one kind of orderly expansion---that dictated by the FLRW
spacetime, at a rate determined by $H(z)$, whose primacy results from
the magnitude of the initially separated energies. In this
view, it is therefore not the behaviour of the various constituents
that determines the expansion profile, as one finds in $\Lambda$CDM but,
rather, the expansion rate is mandated by the Hubble constant and
the properties of FLRW encoded within GR. It is the partitioning of the
various constituents that must follow suit. A useful analogy here is the
chemistry imposed on various interacting elements inside a piston,
whose externally driven evolution in pressure and volume adjusts the
equilibrium densities of the interior components. In other words,
the dynamical behaviour of all the constituents in the cosmic fluid,
including dark energy, is driven by the orderly FLRW expansion, with
a rate based on the initial magnitude of the separated energies,
requiring a fixed $\rho+3p=0$ equation of state.

Future work with the $R_{\rm h}=ct$ hypothesis still has much ground
to cover. For example, the arguments we have made in this paper make
a compelling case for the viability of this model, yet there is one
more type of observation that supersedes all the rest: the actual
measurement of the cosmic expansion rate as a function of redshift
and/or time. In this vein, the so-called `redshift drift' of sources
moving in the Hubble flow has been recognized as a potentially powerful
probe of the background cosmology. Estimates have shown that the first
and second order redshift derivatives can be measured with upcoming
surveys using ELT-HIRES \cite{Liske:2008} and the SKA Phase 2 array
\cite{Klockner:2015}. Even without a detailed calculation, one can
intuitively see that an unambiguous prediction of the $R_{\rm h}=ct$
hypothesis is {\sl zero} drift at all redshifts (see figure~9). This
contrasts sharply with cosmologies, such as $\Lambda$CDM, that predict
a variable expansion rate.

Multi-year monitoring of objects at $z=5$ with the ELT-HIRES will
show a velocity shift $\Delta v = -15$ cm s$^{-1}$ yr$^{-1}$ from
the redshift drift in {\it Planck} $\Lambda$CDM. By comparison, one
expects $\Delta v=0$ cm s$^{-1}$ yr$^{-1}$ in $R_{\rm h}=ct$. The
expected ELT-HIRES measurement error is $\pm 5$ cm s$^{-1}$ yr$^{-1}$
after 5 years of monitoring. Therefore, these upcoming redshift drift
measurements will differentiate between $R_{\rm h}=ct$ and {\it Planck}
$\Lambda$CDM at a confidence level exceeding $3\sigma$, so long as any
possible source evolution is well understood. This is, of course, an
important caveat. Assuming this hurdle can be overcome, such a result
will provide the strongest evidence yet in favour of $R_{\rm h}=ct$.
After $20$ years of monitoring, this program of observations should
favour one of these models relative to the others at better than
$5\sigma$ \cite{Melia:2016c}.

The reason this particular measurement is superior to all the rest is
that, in the end, one does not even need to know the precise value of
$\Delta v$, as long as one can demonstrate without any doubt that it
is not zero. Any drift at all would completely obviate the viability
of the $R_{\rm h}(t)=ct$ hypothesis and re-ignite the quest for
understanding the (most remarkable) $R_{\rm h}(t_0)=ct_0$ coincidence
in cosmology.

\vspace{5mm}
I am very grateful to Amherst College for its support through a John
Woodruff Simpson Lectureship. I am also grateful to the Instituto de
Astrof\'isica de Canarias in Tenerife and to Purple Mountain Observatory
in Nanjing, China for their hospitality while part of this work was carried out.

\end{document}